\documentclass[11pt]{article}

\usepackage{cite}
\usepackage{amssymb}
\usepackage{amsmath}
\usepackage{bm}
\usepackage{graphicx}
\usepackage{braket}

\begin{document}

\title{ Weak Values of Momentum of the Electromagnetic Field: Average Momentum Flow Lines, Not Photon Trajectories.}
\author{R. Flack and B. J. Hiley\footnote{E-mail addresses r.flack@ucl.ac.uk and  b.hiley@ucl.ac.uk.}.}
\date{HEP, University College, Gower Street, \\London WC1E 6BT.\\
\vspace{0.4cm} }
\maketitle

\begin{abstract}
In a recent paper Mahler {\em et al.} have argued that the experiments of Kocsis {\em et al.}  provide experimental evidence for Bohmian mechanics.  Unfortunately these experiments used relativistic, zero rest mass photons whereas Bohmian mechanics is based on non-relativistic Schr\"{o}dinger particles having non-zero rest mass.  The experimental results can be accounted for in terms of a different approach based on the electromagnetic field that was already outlined by Bohm in an appendix of the second of his 1952 papers.  A subsequent development of this approach by Bohm, Hiley, Holland, Kaloyerou, Lam and Dewdney, shows in detail how this theory accounts for experimental quantum phenomena in general.  Using this theory, we are led to the conclusion that the experiments have constructed mean momentum flow lines by measuring the real part of what we term the weak Poynting vector.  These results  support and clarify the analysis of  Bliokh {\em et al}.  The experimental flow lines can be constructed independently of the number of photons in the beam leading to the conclusion that flow lines cannot be interpreted as `photon trajectories'.  We discuss exactly how the notion of a photon arises in the field approach in the light of weak values.

\end{abstract}

\section{Introduction}

 The photon experiment of Kocsis {\em et al.}\cite{ksbbms11} and the recently reported experiment of Mahler {\em et al.} \cite{dmas16} at the University of Toronto have revived the discussion of the meaning of `quantum particle trajectories' particularly in the context of the Bohm interpretation of quantum mechanics~\cite{db52, db52a}.  By measuring the real part of the weak value of the momentum operator, Kocsis {\em et al.}\cite{ksbbms11} construct what they call `trajectories of single photons'. In the second paper, Mahler {\em et al.}~\cite{dmas16} show how the earlier experiments can be used to demonstrate non-local effects, but an essential part of the paper is to argue that their experimental results provide evidence for the correctness of `Bohmian mechanics' including its non-local properties.  
 
 Unfortunately the assumption that the paraxial approximation can be written in a form of the Schr\"{o}dinger equation provided the time $t$ and the mass $m$ satisfy $\hbar t/m=z/k_0$, where $z$ is the position in the longitudinal direction and $k_0$ is the central longitudinal wave vector for the photons, does not justify the use of the non-relativistic Schr\"{o}dinger equation to describe the electromagnetic field~\cite{bbcs13}.
 In this paper we will show that the experiments provide evidence for  Bohm's original proposals for the  treatment of the electromagnetic field and not for a Bohmian mechanics based on the non-relativistic Schr\"{o}dinger particle~\cite{db52a}. (See Bohm, Hiley and Kaloyerou~\cite{dbbhpk87},  Holland~\cite{ph93,ph93a} and Kaloyerou~\cite{pk93, pk06} for further developments.  The numerical simulations of Lam and Dewdney~\cite{mlcd94,mlcd94a} are of particular relevance to the work reported here.) 

It is true that the `trajectories' constructed in the experiments show a similarity to those calculated by Philippidis, Dewdney and Hiley~\cite{cpcdbh79}, but the latter were calculated using the real part of the Schr\"{o}dinger equation under polar decomposition of the wave function for a particle with non-zero rest mass.  If the experiments had used atoms, then a case could have been made for retaining  the notion that atoms follow trajectories even in situations where interference occurs, but even here it cannot be concluded that individual atoms {\em actually follow} these `trajectories'~\cite{db52,db52a,dbbhpk87a,dbbh93,bhdr14}. 


In order to understand the transition to a field approach, we first recall that Bohm made it very clear that although a simultaneous position and momentum can be attributed to the particle, they must be ``regarded as `hidden' since we cannot at present measure them".  Rather than speculating on the precise structure of the quantum particle  which, as he remarked, was ``so little known", he simply asserted that ``one must for all practical purposes restrict oneself to a statistical description of the connection between the values of these variables and the directly observable results of measurement" \cite{db52a}.  Note this was a key assumption and Bohm was aware of its limitations~\cite{db65,db71}.

However even these specific arguments cannot be carried through for photons as Bohm himself had already pointed out~\cite{db52a}.   The photon is a fully relativistic entity with zero rest mass and therefore cannot be described by a non-relativistic Schr\"{o}dinger equation.
 A different treatment is needed for the electromagnetic field.  It is this approach which provides an interpretation of the flow lines determined in the experiments of the Toronto group~\cite{ksbbms11, dmas16}.
 In this paper we will develop the theory and show they are not photon trajectories as suggested.  Although our approach contains the necessary non-local features, we will leave the question of non-locality to a subsequent paper.

   \section{The Bohm Approach to the Electromagnetic Field} \label{sec:BAQE}
  
  \subsection{The Basic Idea}
  
  Since the photon is an excitation of the electromagnetic field, Bohm suggested in his original paper~\cite{db52a} that it was necessary to start, not with the photon itself, but with the electromagnetic field, regarding the field and its conjugate momentum as the analogues of the particle position and the particle momentum in the Schr\"{o}dinger case. 
  
An extension of Bohm's original proposals was made by Bohm, Hiley and Kaloyerou~\cite{dbbhpk87},  confining their attention to the scalar field, showing how these ideas could account for typical interference phenomena such as the two-slit experiment, the photoelectric effect, the  Pfleegor-Mandel experiment~\cite{rplm67} and the non-local EPR effect for entangled `photons' of the scalar field~\cite{dbbh75}.  Lam and Dewdney~\cite{mlcd94,mlcd94a} further clarified these ideas by detailed modelling of cavity scalar field dynamics.  In their first paper~\cite{mlcd94}, by analysing the single-mode and pair-mode cases, they bring out in detail the difference between classical and quantum fields, clearly demonstrating the appearance of non-locality in the quantum case.  Furthermore in their second paper~\cite{mlcd94a}, they demonstrate exactly how particle-like quanta of energy in the boson scalar field appear in this approach without the need for postulating the continual existence of particle-like local bosons.

All of this is carried out for non-relativistic boson fields simply to keep the mathematics as simple as possible so as to present a clear picture of the concepts that are involved. The extension to the more realistic model based on relativistic quantum electrodynamics has been discussed by Holland~\cite{ph93, ph93a} and Kaloyerou~\cite{pk93, pk06}  who applied the field approach to typical electromagnetic interference phenomena such as the two-slit experiment, the photoelectric effect, the  Pfleegor-Mandel experiment~\cite{rplm67} and the non-local EPR effect for entangled photons~\cite{dbbh75}. 
  
  In this approach, it is not the photon that is given a primary role. Rather it is the field itself that is given precedence.  Following on from our experience in the standard approach to quantum field theory, we choose the vector potential $A^\mu(\bm r,t)$, rather than the electromagnetic field itself,  to play the role given to the position variable in the Schr\"{o}dinger approach.  The role of the particle momentum is given to the momentum conjugate to $A^\mu(\bm r,t)$.  As Bell~\cite{jb87} argued, it is the field and its conjugate
momentum that are the `beables' in the electromagnetic theory.  
In order to avoid gauge problems, it is adequate for our discussion of classical fields in this paper to choose the Coulomb gauge $\bm {\nabla.A}=0$.  However in the relativistic case we must choose the Lorentz gauge $\partial  A^\mu/\partial x_\mu=0$.   This does not affect the conclusions of  this paper but for completeness we will comment on this problem in section \ref{sec:4.1}

\subsection{The Classical Electromagnetic Field}  \label{sec:CEF}
  
  To define the context, let us briefly recall the treatment of the classical electromagnetic field. As we have seen, we will need an expression for the momentum conjugate to  $A^\mu(\bm r,t)$. This can be found from  the standard Lagrangian for the classical electromagnetic field\footnote{We put $c=\hbar =1$ throughout this paper.} which is given by
  \begin{eqnarray}
  {\cal {L}}=(\bm E^2-\bm B^2)/2=\frac{1}{2}\left[\left(\frac{\partial \bm A}{\partial t}\right)^2-(\nabla\times \bm A)^2\right]=-F^{\mu\nu}F_{\mu\nu}/4
  \label{eq:Lan}												  \end{eqnarray}
  where $F_{\mu\nu}=\partial_\mu A_\nu-\partial_\nu A_\mu$ is the electromagnetic field tensor.  The conjugate momentum is then
   \begin{eqnarray}
  \Pi_j=\frac{\partial{\cal L}}{\partial \dot A_j}=\frac{\partial A_j}{\partial t}=-E_j.	\label{eq:CCM}
  \end{eqnarray}
  Notice the conjugate momentum is the electric field. We can then write the Hamiltonian as
  \begin{eqnarray}
  {\cal H}=(\bm E^2+\bm B^2)/2=\left[\Pi^2+(\nabla\times \bm A)^2\right]/2.
  \label{eq:Ham}
  \end{eqnarray}
  Finally we note that the Euler-Lagrange equation gives us the usual wave equation
  \begin{equation}
 \nabla^2\bm A-\frac{\partial ^2\bm A}{\partial t^2}=0.	\label{eq:waveeqn}
\end{equation}

 \subsection{The Quantum Field}

In order to quantise the electromagnetic field, the standard approach is to represent the field variables as operators in an abstract Hilbert space.  In the Bohm approach, these variables remain functions on the space-time manifold but the functions $\bm A(x^\mu)$ and $\bm \Pi(x^\mu)$ are subject to a non-commutative product, analogous to the von Neumann-Moyal product introduced in a non-commutative phase space~\cite{bh15}.  
 In other words, we have constructed a generalised non-commutative phase space spanned by the variables $\bm A(x^\mu)$ and $\bm \Pi(x^\mu)$. Rather than developing the Heisenberg algebraic approach along the lines discussed in Hiley~\cite{bh16},
 we will follow Tomonaga~\cite{st46} and introduce a wave functional, $\Psi[\bm A(\bm r, t)]$.  The functional  plays an analogous role to the wave function  in the Bohm particle theory.  For this reason it is called the `super-wave function' in Bohm, Hiley and Kaloyerou~\cite{dbbhpk87}. Its role is to encode the essential features of the overall experimental conditions and it thereby incorporates Bohr's notion of wholeness.    
  
The time evolution of the wave functional follows a Schr\"{o}dinger-like equation
 \begin{eqnarray}
 i\partial_t\Psi=\hat H\Psi	\label{eq:SSchr}
 \end{eqnarray}
 where $\hat H=\int{\cal H}d^3\bm r$ and ${\cal H}(\bm A(\bm r), \bm \Pi(\bm r))$ is the field Hamiltonian, where the operators $\bm A(\bm r)$ and $\bm \Pi(\bm r)$ are independent of time as in the Schr\"{o}dinger picture. We adopt the same interpretation  as Tomonaga and assume the probability of the field configuration, $\bm A'(\bm r)$, is given by,
 \begin{eqnarray}
P[\bm A'(\bm r)]=|\Psi[\bm A'(\bm r)]|^2	\label{eq:prob}
\end{eqnarray}
where  $\bm A'(\bm r)$ is a form of $\bm A(\bm r)$ at a specific time $t'$.

The  separation of time from the space coordinates appears to be against the spirit of relativity.  However the content of the theory is relativistically invariant and we have simply chosen a particular Lorentz frame.  For the purposes of this paper it is sufficient to work in a fixed Lorentz frame as the experiment we are discussing is set up in a specific reference frame, namely the laboratory frame. Therefore it is not necessary to worry about a manifestly covariant formalism although we note that Tomonaga~\cite{st46} has already shown how this can be achieved in his approach.

Continuing with equation (\ref{eq:SSchr}) and substituting the Hamiltonian (\ref{eq:Ham}), we find
 \begin{eqnarray*}
 i\frac{\partial \Psi[\bm A]}{\partial t}=\frac{1}{2}\int d^3\bm r 
 \left(-\frac{\delta^2}{\delta\bm A^2}+(\nabla\times\bm A)^2\right)\Psi[\bm A]
 \end{eqnarray*}
 where $\delta/\delta \bm A$ is the usual notation for a functional derivative. 
 If we now polar decompose the wave functional, $\Psi[\bm A]=R[\bm A]e^{iS[\bm A]}$, and split the resulting equation into its real and imaginary parts, we obtain
  \begin{eqnarray}
 \frac{\partial R^2}{\partial t}+\int d^3\bm r\frac{\delta}{\delta\bm A}\left(R^2\frac{\delta S}{\delta \bm A}\right)=0	\label{eq:CPF}
 \end{eqnarray}
 from the imaginary part.  This equation will be recognised as a continuity equation for $R^2$, the probability defined in equation (\ref{eq:prob}).  The real part gives a quantum Hamilton-Jacobi type equation analogous to the quantum Hamilton-Jacobi equation found in the particle theory.  This now reads
  \begin{eqnarray}
 \frac{\partial S}{\partial t}+\frac{1}{2}\int d^3\bm r\left[\left(\frac{\delta S}{\delta\bm A}\right)^2+(\nabla \times \bm A)^2\right]+Q=0	\label{eq:QHJF}
 \end{eqnarray}
 where
 \begin{eqnarray}
 Q[\bm A]=-\frac{1}{2R}\int d^3\bm r\frac{\delta^2R}{\delta\bm A^2}.	\label{eq:QPF}
 \end{eqnarray}
$ Q[\bm A]$ is the field quantum potential which plays a central role in Bohm's approach~\cite{dbbh93}. 
 Notice that $R[\bm A]$ and $S[\bm A]$ here are functionals and have a role analogous to that played by the two fields $R(x^\mu)$ and $S(x^\mu)$  in the particle theory.  Equations (\ref{eq:CPF}) and (\ref{eq:QHJF}) replace the equations derived from the Schr\"{o}dinger equation for a particle with finite rest mass found in Bohm~\cite{db52a}.  Finally the local field momentum is then given by 
 \begin{eqnarray}
  \Pi(x^\mu)=\frac{\delta S(\bm A(x^\mu))}{\delta \bm A(x^\mu)}		\label{eq:BFMom}
 \end{eqnarray}
 which is the analogue of the local momentum given by $\bm p_B(\bm r,t)=\nabla S(\bm r,t)$ for the Schr\"{o}dinger particle case. By comparing equation (\ref{eq:QHJF}) with the expression for the energy (\ref{eq:Ham}), we find that 
 \begin{eqnarray}
 \bm E=\frac{\delta S}{\delta \bm A}\quad\mbox{and}\quad \bm B= (\nabla\times \bm A)		\label{eq:EandB}
 \end{eqnarray}
 allowing us to express our results in terms of the electric and magnetic fields.
 
Finally we can obtain the equation of motion for the $\bm A(\bm r,t)$ from equation (\ref{eq:QHJF}) by first noting
 \begin{eqnarray*}
 \frac{\delta}{\delta \bm A}\left(\frac{\delta S}{\delta t}\right)=\frac{\delta^2\bm A}{\delta t^2},\quad\quad \frac{\delta}{\delta \bm A(\bm r,t)}\int\left(\frac{\delta S}{\delta \bm A(\bm r',t)}\right)^2d^3\bm r'=0.\\
 \frac{\delta}{\delta\bm A(\bm r,t)}\int[\nabla\times \bm A(\bm r')]^2d^3\bm r'=-\nabla^2\bm A(\bm r,t).\hspace{1.5cm}
 \end{eqnarray*}
 Then the wave equation becomes 
 \begin{eqnarray*}
 \nabla^2\bm A-\frac{\partial ^2\bm A}{\partial t^2}=\frac{\delta Q}{\delta \bm A}
 \end{eqnarray*}
which is the same form as the classical wave equation except for the presence of the quantum field potential.  It is this term that is responsible for producing the quantum effects of the field.

So far we have not mentioned the photon.  Indeed the surprising result that emerges from our approach is that the notion of a photon is not needed to explain the flow lines determined in the experiments.  The fact that this explanation is possible could raise doubts about the validity of the Bohm approach to a quantised  electromagnetic field.  Rest reassured, the Bohm approach does contain the notion of a photon;  it arises from the collective modes of the field. The notion of a photon will be discussed in section \ref{sec:photons}.

\section{The Energy-momentum Tensor and the Weak Values of the Momentum }

\subsection{Particle or field}	\label{sec:porf}

In order to bring out the close similarity between the Bohm particle and Bohm field, let us recall a less well known feature of his approach. The simple step of taking the real and imaginary parts of the Schr\"{o}dinger equation hides a deeper mathematical structure underlying the formalism.  The introduction of a local momentum, in this case the Bohm momentum $\bm p_B(\bm r,t)=\bm\nabla S(\bm r,t)$ has a long history going back to Landau~\cite{ll41}, London~\cite{fl45} and Dirac~\cite{pd45}. (See Berry~\cite{mb13}, Hiley, de Gosson and Dennis~\cite{bhmdggd16} for more detail.)


In this connection Takabayasi~\cite{tt54} had already shown that the local momentum was simply related  to the $T^{0\mu}$ components of the energy-momentum tensor
which, in turn, can be identified with the  weak values of the momentum operator~\cite{bh11}. To show this in detail we must  use the Schr\"{o}dinger Lagrangian introduced by Heisenberg~\cite{wh49}
\begin{eqnarray*}
{\cal L}= - \frac{1}{2m}\nabla \psi^{*}\cdot\nabla \psi+\frac{i}{2}[(\partial_{t}\psi)\psi^{*}-(\partial_{t} \psi^{*})\psi]-V\psi^{*}\psi.	\label{eq:Lagrangian}
\end{eqnarray*}
Application of the Euler-Lagrange equations immediately gives us the Schr\"{o}dinger equation and its dual.  Using this Lagrangian, we find the energy-momentum tensor becomes
\begin{eqnarray}
T^{\mu\nu}=- \left\{\frac{\partial {\cal{L}}}{\partial(\partial^{\mu}\psi)}\partial^{\nu}\psi+\frac{\partial {\cal{L}}}{\partial(\partial^{\mu}\psi^{*})}\partial^{\nu}\psi^{*}\right\}-{\cal L}\delta^{\mu\nu}.	\label{eq:EM}
\end{eqnarray}
The components
\begin{eqnarray*}
T^{0j}=-\left\{\frac{\partial {\cal{L}}}{\partial(\partial^{0}\psi)}\partial^{j}\psi+\frac{\partial {\cal{L}}}{\partial(\partial^{0}\psi^{*})}\partial^{j}\psi^{*}\right\}
\end{eqnarray*}
give the momentum density, which can be reduced to
\begin{eqnarray*}
T^{0j}=\frac{i}{2}\left[\psi^{*}\partial^{j}\psi - \psi\partial^{j}\psi^{*}\right].
\end{eqnarray*}
Writing the wave function in polar form, this becomes
\begin{eqnarray}
T^{0j}=-\rho\partial^jS= \rho\nabla S,      \label{eq:BohmM}
\end{eqnarray}
showing that the Bohm momentum is simply related to $T^{0j}$ through the relation
\begin{eqnarray*}
p^j_B(x,t)=T^{0j}(x,t)/\rho(x,t).
\end{eqnarray*}
Similarly we can show 
\begin{eqnarray*}
T^{00}=-\rho \partial_t S
\end{eqnarray*}
which is the total energy density of the field. Then the Bohm energy is simply
\begin{eqnarray*}
E_B(x,t)=T^{00}(x,t)/\rho(x,t).
\end{eqnarray*}

Although this is presented simply as an illustration here, a similar procedure also works for the Pauli and Dirac fields~\cite{bhbc12} suggesting it has general applicability.

  \subsection{The Energy-momentum Tensor of the Classical EM Field}\label{sec:CEMF}
  
 Let us first recall the details of the energy-momentum tensor for the classical electromagnetic field discussed in subsection \ref{sec:CEF}.   Using the Lagrangian (\ref{eq:Lan}), the energy-momentum tensor becomes
  \begin{eqnarray}
  T^{\mu\nu}=\left(\frac{\partial A^\sigma}{\partial x_\mu}-\frac{\partial A^\mu}{\partial x_\nu}\right)\frac{\partial{\cal L}}{\partial\left(\frac{\partial A^\sigma}{\partial x^\mu}\right)}-{\cal L}g^{\mu\nu}	,	\label{eq:emf}
  \end{eqnarray}
 which can also be written in the form
  \begin{eqnarray*}
  T^{\mu\nu}=-F^{\mu\sigma}F^\nu_\sigma+\frac{1}{4}g^{\mu\nu}F_{\alpha\sigma}F^{\alpha\sigma}.
  \end{eqnarray*}
  In terms of field quantities, this becomes
  \begin{eqnarray*}
  T^{\mu\nu}=E^\mu E^\nu+B^\mu B^\nu-\frac{1}{2}(E^2+B^2)g^{\mu\nu}.
  \end{eqnarray*}
  Defining $S^{\mu\nu}=-T^{\mu\nu}$, we find the momentum density is
  \begin{eqnarray*}
  S^{0j}=({E\times B})^j
  \end{eqnarray*}
  which, of course, is just the Poynting vector.   The electromagnetic energy density is
  \begin{eqnarray*}
  S^{00}=W=-\frac{1}{2}(E^2+B^2).
  \end{eqnarray*}

 \subsection{ The Weak Value of the Field Momentum}
 \label{sec:spinorform}
 
 We now move on to consider the weak value introduced by Aharonov, Albert and Vaidman~\cite{yadalv88}
\begin{eqnarray*}
\langle\hat A_{\psi,\phi}\rangle_w=\frac{\langle\phi|\hat A|\psi\rangle}{\langle\phi|\psi\rangle}.
\end{eqnarray*}
 Flack and Hiley~\cite{bh11} have shown that the real part of the weak value of the momentum defined by 
 \begin{eqnarray*}
 \langle \hat P_{\psi,x}\rangle_w=\frac{\braket{ x|\hat P|\psi}}{\braket {x|\psi}}
 \end{eqnarray*}
corresponds to the local or Bohm momentum which was, in turn, related to the $T^{0j}$ component of the energy-momentum tensor through the relation (\ref{eq:BohmM}). Following a similar line of argument, Holland~\cite{ph93} had earlier introduced the notion of a local expectation value of a general operator $\hat A$ for a scalar field defined by 
\begin{eqnarray*}
A[\psi]=Re\left[\Psi^*[\psi](\hat A\Psi[\psi])/|\Psi[\psi]|^2\right]
\end{eqnarray*}
This becomes the real part of the weak value of the field momentum when we choose $\hat A=\widehat T^{\mu\nu}$ and replace the the scalar field by the vector potential.  The real part of the weak value of the field momentum density operator  then becomes
 \begin{eqnarray}
 \langle \widehat T^{\mu\nu}_{\Psi,\bm A}\rangle_w=\frac{\langle\bm A|\widehat T^{\mu\nu}|\Psi\rangle}{\langle\bm A|\Psi\rangle}	\label{eq:wvf}
 \end{eqnarray}
 where we will take the momentum density operator in this case to be
 \begin{eqnarray}
 \widehat T^{0j}=\frac{i}{2}\left((\nabla\times \bm A)^j\frac{\delta}{\delta \bm A}+\frac{\delta}{\delta\bm  A}(\nabla\times\bm A)^j\right).	\label{eq:Pop}
 \end{eqnarray}
 The real part of the weak value of the momentum density then becomes 
 \begin{eqnarray}
  \langle \hat T^{0j}_{\Psi,\bm A}\rangle_w=-[\delta S/\delta\bm A\times(\nabla\times \bm A)]^{j}= (\bm E\times\bm B)^j, \label{eq:wvt}
 \end{eqnarray}
   and we interpret this as the real part of the {\em weak value}  of  the Poynting vector which, in our approach, is complex.  The real part of this vector describes the flow of energy-momentum at each point of space-time.   
 We also find the local energy density  $\langle T^{00}_{\Psi,\bm A}\rangle_w(x^\mu)$ is given by
 \begin{eqnarray}
 \langle T^{00}_{\Psi,\bm A}\rangle_w(x^\mu)=\frac{1}{2}\left[(\delta S/\delta \bm A)^2+ (\nabla\times \bm A)^2 \right]+Q_{\bm A}=\frac{1}{2}	[E^2+B^2]+Q_{\bm A},	\label{eq:TFEnerg}
 \end{eqnarray}

where
\begin{eqnarray}
 Q_{\bm A}=-(1/2R)\delta^2R/\delta\bm A^2.	\label{eq:QA}
 \end{eqnarray}
Equations (17)-(19) are a straightforward generalisation of the scalar field equations (5.29)-(5.31) appearing in Holland~\cite{ph93}

 From these equations we immediately see that the local energy density is made up of two parts, the local electromagnetic energy density plus the quantum potential energy density.  Hence, just as in the case of a particle with finite rest mass, some of the energy that is available classically for the em field, is taken up by the quantum potential. The above discussion is concerned with the real part of the weak value.  We will discuss the imaginary part later.

The existence of the weak value should enable us to determine energy-momentum flow lines of the type found by  Kocsis {\em et al.}\cite{ksbbms11} and by Mahler {\em et al.} \cite{dmas16}.  To bring out the exact meaning of these flow lines in more detail, we first recall some well-known properties of the classical energy-momentum tensor $T^{\mu\nu}$.  Firstly it provides a measure of the total four-momentum crossing an element of area $dS$ which is given by $T_{\mu\nu}N^\mu dS$ where $N^\mu$ is a normal to $dS$. 

Secondly we know that for every pair of future-pointing causal vectors $U^\mu$ and $V^\nu$, $T_{\mu\nu}U^\mu V^\nu\geqslant 0$~\cite{rpwr84}.  This means that if $V^\mu$ is the 4-velocity of an observer, then we can identify  $V^\nu T^\mu_\nu$ with the Poynting 4-vector.  We can therefore define a current $j^\mu= V^\nu T^\mu_\nu $ with $V^\mu$ a future pointing vector with $V^\mu V_\mu \geqslant 0$ and $V^0>0 $ so that 
 \begin{eqnarray*}
 j^\mu j_\mu\geqslant0,\quad\mbox{and}\quad V^\mu j_\mu\geqslant0.
 \end{eqnarray*}
 Thus $j^\mu$ is a future pointing, time-like or null 
 vector.  Furthermore if $V^\mu$ is a constant vector, $j^\mu$ is conserved so that $\partial_\mu j^\mu=0$.  Hence we can interpret $j^\mu$ as the electromagnetic energy-momentum density four-vector with the spatial components $\bm j$ giving the momentum density.  With these variables we are able to track the energy flow through the equation
 \begin{eqnarray*}
 dx^k/dt=(j^k/j^0)|_{\bm x=\bm x(t)}=(V_\nu T^{\nu k}/V_\nu T^{\nu0})|_{\bm x=\bm x(t)}.
 \end{eqnarray*}
 If we now choose $V^\mu=\delta^\mu_0$ this equation becomes
 \begin{eqnarray}
 \dot x^k=T^{0k}/T^{00}|_{\bm x=\bm x(t)}	\label{eq:cflowline}
 \end{eqnarray}
 which is just the Poynting vector divided by the mean energy density. So it is clear that classically, equation (\ref{eq:cflowline}) gives an energy-momentum flow line,  which is what we would expect.

\subsection{The Quantum Field}

When we turn to consider the quantum case, care must be taken to avoid the infinities that plague field approaches.  The first step is to remove the infinity due to the infinite energy of the ground state.  This is done by  adopting  the usual procedure of normal ordering~\cite{ss64,ph93} without affecting the mean momentum density.  It is then possible to define a modified current 
\begin{eqnarray*}
j^\mu=V_\nu \langle\bm A|:\widehat T^{\mu\nu}:|\Psi\rangle=V^\nu\langle\widehat T^{\mu\nu}_N\rangle_w.
\end{eqnarray*}
Since $V^\mu$ is a future pointing vector,  it can be shown that, in the case of Fock and coherent states,  $j^\mu$ is a conserved, future pointing  vector.  Thus we can re-write equation (\ref{eq:cflowline}) in the form
\begin{eqnarray*}
\dot x^j=\langle \widehat T^{0k}_N\rangle_w/\langle\widehat T^{00}_N\rangle_w|_{x=x(t)}.
\end{eqnarray*}
It is these flow lines that have been constructed in the Toronto experiments.  
 
 The experiments were performed using a very weak source comprising a quantum dot and this implies that  there was only one photon present in the apparatus at a given time.  In our discussion so far there is no evidence of any photon-like entity in  the quantum field governed by the equations (\ref{eq:CPF}) and (\ref{eq:QHJF}). In fact the momentum flow lines are independent  of the specific number of photons present in the beam.  This should not come as too much of a surprise because, as we have already remarked, our results would also apply to coherent states where the number of photons is not well defined.  Indeed our results predict that the flow lines are independent of the photon content of the field. 
 
 The fact that in the case of fields, these flow lines cannot be attributed to individual photons, or boson quanta has been pointed out many times in the papers of Bohm, Hiley and Kaloyerou~\cite{dbbhpk87},  Holland~\cite{ph93, ph93a},  Kaloyerou~\cite{pk93, pk06} and Lam and Dewdney~\cite{mlcd94,mlcd94a}.
 So we are left with the question as to exactly how the notion of a photon arises in this interpretation.  We will leave that question aside for the moment, but will return to it in section \ref{sec:photons}.

\section{Experimental Determination of   $ \langle T_{\Psi,\bm A}^{0k}\rangle_w$}

\subsection{ The Role of the Thin Calcite Crystal}  \label{sec:4.1}

Let us now apply the Bohm theory of the electromagnetic field to the actual Toronto experiments.  As shown in Duck, Stevenson and Sudarshan~\cite{idpsgs89} the
crucial step in determining a weak value in the particle case is to look for a coupling between the centre of mass wave function and the spin.  In other words, we need to introduce a coupling between the momentum of the particle and its spin.  In our case we then require a coupling between the field momentum and its polarisation.  Such a process can be physically realised  in the phenomenon of birefringence.  In a uniaxial crystal, we have a difference of momentum between the ordinary and extraordinary rays;  the ordinary ray being polarised perpendicular to the optic axis, while the extraordinary ray is polarised along the optic axis.

To see how this comes in, let us recall that, for a ray travelling along the $z$-axis, the paraxial approximation allows us to write the electric field in the following form
\begin{eqnarray*}
\bm E(\bm r,t)=\bm E(x,y)e^{i(kz-\omega t)}
\end{eqnarray*}
with
\begin{eqnarray}
\bm E(x,y)=\begin{pmatrix}e^{i\delta_1/2}E_x\\e^{i\delta_2/2}E_y\end{pmatrix}	\label{eq:Exy}
\end{eqnarray}
where $\delta_2-\delta_1$ is the phase difference introduced as the ray travels through the medium.  Thus in effect, we have a two state system because the electromagnetic field has two eigenstates of polarisation, $\pm 1$, corresponding to 
left and right circular polarisation.
In terms of the vector potential this expression becomes
\begin{eqnarray}
\bm A(x,y)=\begin{pmatrix}e^{i\delta_1/2}A_x\\e^{i\delta_2/2}A_y\end{pmatrix}= \begin{pmatrix} 
      \zeta_x \\\zeta_y \end{pmatrix}.	\label{eq:Axy}
\end{eqnarray}

At this point there is one more comment to make, namely, when we are dealing with a relativistic theory we must look closely at the gauge condition.  As is well known, in quantum electro-dynamics we must remove the longitudinal and time-like part of the vector potential $A^\mu$.  We do this by imposing the Gupta-Bleuler subsidiary condition $\partial_\mu A^{(+)}_\mu|\Psi\rangle=0$, where $A^{(+)}_\mu$ is the destruction part of the operator~\cite{ss64a}.  This is sufficient to ensure the field remains transversally polarised.  Thus our argument goes through in any reference frame.

Equation (\ref{eq:Axy}) takes care of the polarisation aspects of the vector potential, but makes no provision for the width of the pulse of radiation used in the Toronto experiments.  For simplicity we will assume that we are dealing with a Gaussian pulse with a circular cross-section of width $q(z)$.  Then the vector potential is 
\begin{eqnarray}
\bm A(\bm r)= \frac{1}{q(z)}\exp\left(-ik\frac{r^2}{2q(z)}\right)   \begin{pmatrix} 
      \zeta_x \\\zeta_y \end{pmatrix}
\quad\mbox {with}\quad r^2=x^2+y^2	\label{eq:Exyspinor}
\end{eqnarray}
where the two polarisation states of the vector potential have been written as a column matrix, indicating that the polarisation can be written as a two-component spinor.

We are now in a position to analyse what happens as a polarised beam enters a thin calcite crystal placed in the interference region of the two-slit setup used by Kocsis {\em et al.}~\cite{ksbbms11}.  As the beam passes through the uniaxial crystal at an angle to the optic axis, the two components of the polarisation move with different speeds causing the polarisation vector to rotate.  This rotation can be analysed by introducing the interaction Hamiltonian $\widehat {H_I}=g {\widehat T^{0j}\hat S_j}$ where $\widehat T^{0j}$ is given in equation (\ref{eq:Pop}).  If we write the wave functional of the field $\Psi[\bm A(\bm r)]$ more formally as 
\begin{eqnarray*}
|\Psi_i\rangle=|\psi\rangle|\zeta\rangle,
\end{eqnarray*}
then the state after the beam leaves the crystal is
\begin{eqnarray}
|\Psi_f\rangle=e^{-i\widehat {H_I}t}|\Psi_i\rangle.		\label{eq:Inteqn}
\end{eqnarray}
Kocsis {\em et al.}~\cite{ksbbms11} were interested in the weak value of the transverse momentum in the interference region, so using the paraxial approximation in which the beam travels along the $z$-axis, equation (\ref{eq:Inteqn}) becomes
\begin{eqnarray}
|\Psi_f\rangle=e^{-ig\widehat T^{0x}_N.\hat S_xt}|\psi\rangle|\zeta'\rangle.	\label{eq:AA}
\end{eqnarray}

Using the operator (\ref{eq:Pop}), together with  $\hat S_x=(|H\rangle\langle H|-|V\rangle\langle V|)/2$, which is the polarisation operator used in the Toronto experiment, in equation (\ref{eq:wvf}), with the initial state chosen to be $|\zeta'\rangle=(|H\rangle+|V\rangle)/\sqrt{2}$, it is straightforward to show that
\begin{eqnarray}
\langle x_f|\Psi_f\rangle\approx \frac{\psi(x_f)}{\sqrt{2}}\left(e^{-\frac{igt}{2} \langle \widehat T^{0x}_N\rangle_w}|H\rangle+e^{\frac{igt}{2}\langle \widehat T^{0x}_N\rangle_w}|V\rangle\right)		\label{eq:finalresult}
\end{eqnarray}
provided the crystal is thin enough.  Thus the crystal produces a rotation in the polarisation vector which depends upon the weak value of the transverse momentum. As we have shown in subsection \ref{sec:spinorform}, this weak value is just the $x$-component of the weak value of the normal ordered expectation value of Poynting vector defined in equation (\ref{eq:wvt}).

\subsection{The Real Part of the Weak Value}

This result then leads directly to that obtained by Kocsis {\em et al.}~\cite{ksbbms11} with the real part of their $\langle \hat k_x\rangle_w$ replaced by the real part of $\langle T^{0x}_N\rangle_w$.  Writing $\hat S_z=|R\rangle\langle R|-|L\rangle\langle L|$, we can now find
\begin{eqnarray*}
\langle \hat S_z\rangle=\frac{I_R-I_L}{I_R+I_L}.
\end{eqnarray*}
By measuring the intensities of the right- and left-handed components of the final polarisation states of the field,  we are able to find the real part of the weak value of the transverse momentum of the field through the relation
\begin{eqnarray*}
\Re\langle T^{0x}_N\rangle_w=\xi \sin^{-1}\left(\frac{I_R-I_L}{I_R+I_L}\right)
\end{eqnarray*}
where $\xi=\frac{|\bm k|}{\zeta}$ is determined in  the experiment itself.  This is exactly the result that Kocsis {\em et al.}~\cite{ksbbms11} have measured.  These results were then used to construct flow lines.  Thus our approach gives a different interpretation to these flow lines.  We will discuss the differences in more detail below.

\subsection{The Imaginary Part of the Weak Value}

It is also possible to obtain values for the imaginary part of $\langle T^{0x}_N\rangle_w$.  This is done by putting the imaginary part into equation (\ref{eq:finalresult}) obtaining
\begin{eqnarray}
\langle x_f|\Psi_f\rangle\approx \frac{\psi(x_f)}{\sqrt{2}}\left(e^{\frac{gt}{2}\Im\langle \widehat T^{0x}_{N}\rangle_w}|H\rangle+e^{-\frac{gt}{2}\Im\langle \widehat T^{0x}_{N}\rangle_w}|V\rangle\right).		\label{eq:imagfinal}
\end{eqnarray}
Then using $\hat S_x=(|H\rangle\langle H|-|V\rangle\langle V|)/2$ we find 
\begin{eqnarray*}
\langle \hat S_x\rangle=\frac{I_H-I_V}{I_H+I_V}.
\end{eqnarray*}
This then gives
\begin{eqnarray*}
\Im\langle T^{0x}_N\rangle_w=\xi \sinh^{-1}\left(\frac{I_H-I_V}{I_H+I_V}\right).
\end{eqnarray*}

If one evaluates the mean value of $\hat S_y=|H\rangle\langle V|+|V\rangle\langle H|$, we find $\langle \hat S_y\rangle=1$.  The imaginary parts simply mean that the polarisation has changed from circular to elliptical polarisation.  Thus the imaginary part $\Im\langle T^{0x}_N\rangle_w$ measures the degree of absorption of the light ray passing through the calcite crystal.

\subsection{Discussion of Results}

Note, these result have been obtained without reference to the notion of a photon as a particle.  Secondly  equation (\ref{eq:wvt}) shows we are measuring the real part of the {\em weak value} of the Poynting vector. The flow lines are momentum flow lines of the field and are independent of the  field's photon content.  

At first sight this seems a surprising conclusion but it actually explains the results obtained by Bliokh {\em et al.}~\cite{kbab13}. 
Using a purely classical argument based on the behaviour of electric fields in uniaxial crystals, they show how exactly the same results can be predicted. But, of course, this does not explain the complex values which are a consequence of the quantum formalism.  However it is now clear why the approach of Bliokh {\em et al.} works.  

 Both the electric and magnetic fields are weak values because the calcite crystal is in the interference region.  In that region there exists quantum potential energy as equation (\ref{eq:TFEnerg}) clearly shows. The presence of the quantum potential energy changes the amount of electromagnetic energy present and thus the field strengths must be different from what they would have been had the calcite crystal not been in the interference region.  Indeed if these variables had their classical values there would be no interference and the momentum flow lines would radiate out radially.  Thus we see exactly how  the quantum potential accounts for the form of the flow lines.  One major question remains, ``How does the concept of a photon arise in this version of the Bohm approach?"

\section{Photons in the Bohm Approach}	\label{sec:photons} 

 \subsection{The Normal Modes of the Electromagnetic Field} \label{sec:cp}  
  
Finally to allay doubts concerning the quantum nature of our approach, we will now discuss in detail how the notion of a photon arises. To show this we go to what is essentially the momentum representation, since the momentum of the photon is well defined.   This means we must  expand $\bm A(x^\mu)$ in terms  of normal modes, which requires the introduction of a two component vector $q_{k,\mu}$, orthogonal to $\bm k$.  Thus we can write
  \begin{eqnarray}
   {\bm A}(\bm r,t)=(1/V)^{1/2}\sum_{k,\mu}\epsilon_{k,\mu}q_{k,\mu}(t)e^{i\bm{k.r}}.		\label{eq:WsumNM}
  \end{eqnarray}
Here $\bm k$ is the wave number of the normal mode and  $\epsilon_{k,\mu}$ is  the polarisation vector with components, $\mu=1,2$. $V$ is the volume of the box, which is assumed large.   
 The corresponding conjugate momentum expanded in terms of the $\pi_{k,\mu}(t)$ is given by
  \begin{eqnarray}
 \bm \Pi(\bm r,t)=(1/V)^{1/2}\sum_{k,\mu}\epsilon_{k,\mu}\pi_{k,\mu}(t)e^{-i\bm{k.r}}.		\label{eq:gcm}
  \end{eqnarray}
  
  The set of variables $q_{k \mu}$ and $\pi_{k \mu}$  describe the normal modes of the field and they become our beables in this representation.
Since the vector potential and its conjugate momentum are real, we can write $q_{k,\mu}=q^\dag_{-k,\mu}$ and $\pi_{k,\mu}=\pi^\dag_{-k,\mu}$.  We can regard these variables as spanning a  phase space since each pair satisfy the Poisson bracket relationships~\cite{dbbhpk87}
\begin{eqnarray*}
    \{\pi_{k,\mu},q_{k',\mu'}\}=\{\pi^\dag_{k,\mu},q^\dag_{k',\mu'}\}=-i\delta_{kk'}\delta_{\mu\mu'}.
\end{eqnarray*}
Clearly equations (\ref{eq:WsumNM}) and  (\ref{eq:gcm}) involve many modes and so this is analogous to treating a many particle situation in the non-relativistic case. 

Now  we introduce a wave functional of all the normal modes $\Psi(\dots q_{k,\mu}\dots q^\dag_{k,\mu}\dots) $.  The corresponding Hamiltonian (\ref{eq:Ham})  of the system becomes 
  \begin{eqnarray*}
  H=\sum_{k,\mu}\left(\pi_{k,\mu}\pi^*_{k,\mu}+\kappa^2c^2q_{k,\mu}q^*_{k,\mu}\right)
  \end{eqnarray*}
  where $\kappa=|\bm k|$.  This shows that the field can be described as a set of  harmonic oscillators.  If we now substitute this Hamiltonian into the super-Schr\"{o}dinger equation (\ref{eq:SSchr}), using the super-wave function
  \begin{eqnarray*}
      \Psi(\dots q^*_{k,\mu}\dots q_{k,\mu}\dots)=R(\dots q^*_{k,\mu}\dots q_{k,\mu}\dots)\exp[iS(\dots q^*_{k,\mu}\dots q_{k,\mu}\dots)], 
  \end{eqnarray*}
 we arrive at the quantum Hamilton-Jacobi equation \begin{eqnarray}
\frac{\partial S}{\partial t}+\sum_{k,\mu}\frac{\partial S}{\partial q_{k,\mu}}
\frac{\partial S}{\partial q_{k,\mu}^*}+\sum_{k,\mu}\kappa^2q_{k,\mu}q_{k,\mu}^*+Q(\dots q^*_{k,\mu}\dots q_{k,\mu}\dots)=0,		\label{eq:QHJquanta}
\end{eqnarray}
which is the equation of motion of a set of  quantum harmonic oscillators. If the set of harmonic oscillators are independent then the equation of motion of each one can be written in a more familiar form
\begin{eqnarray}
\ddot{q}_{k,\mu}+\kappa^2q_{k,\mu}=\frac{\partial}{\partial q_{k,\mu}}
\left(\frac{1}{2R}\sum_{k',\mu'}\frac{\partial^2R}{\partial q_{k',\mu'}\partial q_{k',\mu'}^*}   \right)	.	\label{eq:QHO}
\end{eqnarray}
As we will show in the next section, it is these quantised modes that enable us to talk about photons. Notice that  equation (\ref{eq:QHJquanta}) contains an expression for the quantum potential, $Q(q^*_{k \mu},q_{k \mu})$ which, for an independent mode,  takes  the form
\begin{eqnarray*}
Q=-\sum_{k,\mu}\frac{1}{2R}\frac{\partial^2R}{\partial q^*_{k,\mu}\partial q_{k,\mu}}.
\end{eqnarray*}
Note that if the quantum potential cannot be factorised into single mode variables, then the oscillators are said to be entangled.  This then becomes the source of the non-locality discussed in the experiment of Mahler {\em et al.} \cite{dmas16}.

For completeness we write down the corresponding continuity equation \begin{eqnarray}
\frac{\partial R^2}{\partial t}+\frac{1}{2}\sum_{k,\mu}\left[\frac{\partial}{\partial q_{k,\mu}}\left(R^2\frac{\partial S}{\partial q^*_{k\mu}}\right)+	\frac{\partial}{\partial q^*_{k,\mu}}\left(R^2\frac{\partial S}{\partial q_{k\mu}}\right)	\right]
\end{eqnarray}
describing the conservation of probability.

\subsection{The Energy-momentum Tensor: Normal Modes}

To see how the notion of a photon arises in the source used in the experiment of Kocsis {\em et al.}\cite{ksbbms11}, we need to examine the expression for the energy density
 \begin{eqnarray}
{\cal E}(\bm r)=\frac{1}{2V}\sum_{k,\mu}\sum_{k',\mu'}\left[\frac{\partial S}{\partial q^*_{k',\mu'}}\frac{\partial S}{\partial q_{k,\mu}}+2Q	\right]\epsilon_{k,\mu}\epsilon_{k',\mu'}e^{-i\bm{(k-k').r}}\nonumber\\+
(k\times\epsilon_{k,\mu})(k'\times\epsilon_{k',\mu'})q^*_{k',\mu'}q_{k,\mu}e^{i\bm{(k-k').r}}		\label{eq:Eden}
 \end{eqnarray}
where $(k\times\epsilon_{k,\mu})$ is a vector product.  The momentum density is given by
 \begin{eqnarray}
 \bm S(\bm r)=-\frac{i}{V}\sum_{k,\mu}\sum_{k',\mu'}\left[(\epsilon_{k',\mu'}\times (k\times\epsilon_{k,\mu}))\frac{\partial S}{\partial q_{k',\mu'}}q_{k,\mu}e^{i\bm {(k-k').r}}\right].		\label{eq:Mden}
 \end{eqnarray}
 The derivation will be found in Holland~\cite{ph93} and in Kaloyerou\cite{pk93, pk06}. 
 
 \subsection{The Energy Content of the State Containing One Photon}

To obtain a clearer picture of how the photon arises in the quantum state used in the experiments of Kocsis {\em et al.}\cite{ksbbms11}, let us explore the behaviour of the momentum density for the ground state and the first excited state.  
 
 The ground state wave function is given by
 \begin{eqnarray*}
 \Phi_0=\exp\left[-\sum_{k,\mu}\frac{\kappa}{2}q^*_{k,\mu}q_{k,\mu}\right]\exp\left[-\sum_{k,\mu}\frac{i\kappa t}{2}\right ]
 \end{eqnarray*}
 which gives
 \begin{eqnarray*}
 R=\exp\left[-\sum_{k,\mu}\frac{\kappa}{2}q^*_{k,\mu}q_{k,\mu}\right]\quad\mbox{and}\quad S=-\sum_{k,\mu}\frac{\kappa t}{2}.
 \end{eqnarray*}
 Here $\kappa=|\bm k|$.
 Since $S$ is not a function of $q_{k,\mu}$, the momentum density is zero which is the typical result expected for the ground state.  The total energy is 
 \begin{eqnarray*}
 E=-\frac{\partial S}{\partial t}=\sum_{k,\mu}\frac{\kappa}{2},
 \end{eqnarray*}
which is just the usual expectation value of the ground state energy -- the zero-point energy.
It is interesting to note that the quantum Hamilton-Jacobi field equation (\ref{eq:QHJquanta}) is 
\begin{eqnarray}
-\frac{\partial S}{\partial t}=\sum_{k,\mu}\frac{\kappa^2}{2}q^*_{k,\mu}q_{k,\mu} +Q.		\label{eq:gse}
\end{eqnarray}
The field quantum potential is
\begin{eqnarray*}
Q=-\sum_{k,\mu}\frac{\kappa^2}{2}q^*_{k,\mu}q_{k,\mu}+\sum_{k,\mu}\frac{\kappa}{2}.
\end{eqnarray*}
Comparing this with equation (\ref{eq:gse}), we see that the zero-point energy is contained in the field quantum potential.  However as we have remarked above this energy is infinite and this must be removed by normal ordering.

In the experiment of Kocsis {\em et al.}~\cite{ksbbms11}, the source of the electromagnetic field produced only one photon at a time.  To understand this situation even though it has no relevance for the momentum flow lines, we will for completeness consider what happens in the case of the first excited state of the field.  This is given by
\begin{eqnarray*}
\Phi_{k,\mu}=(2\kappa)^{\frac{1}{2}}q^*_{k,\mu}\Phi_0\exp[-i\kappa t].
\end{eqnarray*}
The momentum density in this state is
\begin{eqnarray}
\bm S(\bm r)=-\frac{1}{2V}\sum_{k,\mu}\left[	[\epsilon_{k',\mu'}\times(k\times\epsilon_{k,\mu})]\frac {q_{k,\mu}}{(-1)^{\mu'}q^*_{k',\mu'}}e^{i\bm{(k+k').r}}	\right]\hspace{1cm}\nonumber\\+-\frac{1}{2V}\sum_{k,\mu}\left[	[\epsilon_{k',\mu'}\times(k\times\epsilon_{k,\mu})]\frac {q_{k,\mu}}{q_{k',\mu'}}e^{i\bm{(k-k').r}}	\right].	\label{eq:localmom}
\end{eqnarray}
Notice here that the momentum density is a function of not one excitation, but a linear combination of excitations.  What this means is that there is no way to separate out the space-time properties of a single photon.  So what, then, do we mean by a single photon state? We mean that the total momentum of the state is just $\bm k'$.  Indeed the total momentum of the state is given by the integration of equation (\ref{eq:localmom}) over all space, i.e.,
\begin{eqnarray*}
\bm S=\int \bm S(\bm r)d\bm r=\bm k'
\end{eqnarray*}
and the energy
\begin{eqnarray*}
E=\sum_{k,\mu}\frac{\kappa}{2}+\kappa'
\end{eqnarray*}
showing that the field contains one quantum of energy, $h\nu$, above the zero-point energy and one quantum of momentum $\hbar k'$. This is the energy and momentum of the photon.

In practice it is not possible to have a source that produces photons with the same sharply defined wave number, $\bm k'$, so we  assume that the state will produce a small distribution of wave numbers, $f_\mu(\bm k-\bm k')$,  which will be sharply peaked around a mean value $\bm k'$.  The state we are analysing is now described by
\begin{eqnarray*}
\Phi_p=\sum_{k,\mu}f_\mu(\bm k-\bm k')\sqrt{2\kappa}\;q^*_{k,\mu}\Phi_0e^{-i\kappa t}.
\end{eqnarray*}
Thus in a weak beam, we will find the mean value of the momentum of the photons will  given by
\begin{eqnarray*}
\langle \bm S\rangle= \bm k'.
\end{eqnarray*}
This means that in position space, we have a wave packet whose width depends on the factor $f_\mu$. 

There are significant differences in dealing with the photon because we are  in the relativistic region where vacuum fluctuations  play an important role.  In the non-relativistic Bohm particle model,  we assume that we are below the threshold for pair creation and so we are always dealing with single particle states.  However if we generalise the model to encompass Dirac relativistic particles, as has been done by Hiley and Callaghan~\cite{bhbc12},  we must ensure we are dealing with particles that have energies below the pair creation threshold. An interesting discussion of this problem can be found in Kemmer~\cite{nk39}.


\subsection{Emergence of Energy-momentum Flow Lines}

We now have expressions for both the energy density, (\ref{eq:Eden}), and the momentum density, (\ref{eq:Mden}), as a function of position giving us a local expression for the total energy-momentum flow.  Furthermore we have shown that these expressions are simply the weak values of the energy and the momentum $\langle T^{00}_N\rangle_w$ and $\langle T^{0j}_N\rangle_w$ that appear in equations (\ref{eq:TFEnerg}) and (\ref{eq:wvt}).  Note that the weak values depend on the electric and magnetic fields in the region of interference. Thus we have linked the flow lines to the field variables.  

Now let us compare this field approach with the Bohm approach for a single particle.  In the particle case, the Bohm momentum is the real part of the weak value of the momentum and is given by the canonical momentum $\bm P_B=\nabla S$ where $S$ is the phase of the wave function.  For the electromagnetic field, the classical canonical momentum is given in equation (\ref{eq:CCM}), namely, $\Pi_j=\partial {\cal L}/\partial {\bm \dot A}=-E_j $, where $Ej$ is the electric field.  In the quantum case, the canonical momentum is $\Pi_j=\delta S/\delta\bm A$ which appears as $\bm E$ in equations (\ref{eq:TFEnerg}) and (\ref{eq:wvt}). This step is the analogue of replacing the action with the phase in the expression for the local momentum in the particle theory as we move from the classical to the quantum domain.

Here the analogy ends, unless we can localise the photon, but this is just what we cannot do.  As we have seen in section \ref{sec:cp}, the normal modes are characterised by $q_{k,\mu}$ and $\pi_{k,\mu}$ and these expressions are not localised in space-time. They are the parameters of an energy-momentum space.   If we take the inverse Fourier transform of equation (\ref{eq:WsumNM}) we find
\begin{eqnarray*}
q_{k,\mu}=\frac{\epsilon_{k,\mu}}{V^{1/2}}\int \bm A(x^\mu)\exp[-i\bm{k.r}]d^3\bm r.
\end{eqnarray*}
A  corresponding expression for $\pi_{k,\mu}$ arises from taking the inverse Fourier transform of equation (\ref{eq:gcm}).  These parameters are not local and so it is clearly  not possible to give a local description of the photon.  Therefore no meaning can be given to the notion of a `photon trajectory' confirming our conclusion that the experiments have constructed momentum flow lines.

\subsection{The Appearance of the Photon}

How, then, does the notion of a photon arise in this approach?  Here we arrive at the same conclusion as that reached in standard quantum field theory.    Muthukrishnan, Scully and Zubairy~\cite{amms08} 
 conclude that the only way we can localise the photon is via absorption or emission by an atom.  There is no position ket $|x^\mu
  \rangle$ for a photon.  Then we can talk about the probability of absorbing a photon in the region of the atom by 
\begin{eqnarray*}
P(\bm r,t)=\langle 0|E^+(\bm r,t)|\psi\rangle=\kappa|\Psi[\bm A'(\bm r,t)]|^2.
\end{eqnarray*}
A discussion of how this process is accounted for in the Bohm scalar field model is presented in Lam and Dewdney~\cite{mlcd94a}. Again the field quantum is localised by an absorbing atom.  Indeed Dewdney and Lam~\cite{cdml91} explained in detail how this absorption process takes place in an individual atom.

Once this event has taken place, the photon can no longer be said to exist, and the field configuration has changed.  Similarly for the emission process in which a photon is created in the sense that the atom releases a quantum of energy into the field, giving rise to a new field configuration, $\Psi[\bm A(\bm r,t)]$, when the energy is no longer localised at the atom. 

 Bohm, Hiley and Kaloyerou\cite{dbbhpk87} have explained how this process works from the perspective of their approach to the electromagnetic field in the following way.  The essential point is that the non-linear and non-local action of the quantum potential sweeps out the energy from the field and this energy produces an excited state of, say, an atom.  As a result the photon ceases to exist as an entity.  
 
 The whole process was discussed by illustrating what is involved in a simplified model of an Auger-like transition.  Bohm, Hiley and Kaloyerou\cite{dbbhpk87} show in detail how, using an appropriate interaction Hamiltonian between the atom and the electromagnetic field, the energy is swept up through the non-local and non-linear effect of the quantum potential that appears in the real part of the interaction equation. More details of this whole process can be found in Holland~\cite{ph93} and in Kaloyerou\cite{pk93, pk06}.

 Unlike the particle theory, where the classical limit emerges via a smooth transition to the classical particle trajectory~\cite{bham95}, the treatment of the electromagnetic field has no particle-like classical limit.   The classical limit is a {\em field}.  This is emphasised in the standard approach which shows that the classical limit emerges from the coherent state  rather than a Fock state.
In a coherent  state, as opposed to a Fock state, the photon number is not well defined. Nevertheless even in the coherent state there is still a local weak Poynting vector (\ref{eq:wvt}) which, as we have explained, enables us to determine a momentum flow line.

This leaves the question as to how such a process can explain the cloud chamber track left by a $\gamma$-ray.
The process is quite complex depending on the nature of the actual transition process.  Nevertheless, conservation of  momentum ensures that the electromagnetic energy is emitted in a more or less well-defined direction.  Thus, for example, the track of a photon in a tracking device can be explained qualitatively as follows. At each scattering the quantum of energy is swept up into the atom and then there is re-emission of a quantum of slightly lower energy moving in a somewhat different direction. If the initial energy of the quantum is large compared with that absorbed by the atom, then a series of such inelastic scatterings will define something resembling a jagged track.  This is essentially the idea lying behind Mott's explanation of detector trajectories.  A more detailed treatment of this problem in terms of the model we are discussing here has been given by de Gosson, Hiley and Cohen~\cite{mdgbhec16}.

\section{Conclusion}

We have shown how the experiments of Kocsis {\em et al.}~\cite{ksbbms11} arise from the treatment of the electromagnetic field as already outlined in the second of Bohm's seminal 1952 papers~\cite{db52a}.  By extending the approach to discuss the weak value of the field momentum, we arrive at a weak value for the Poynting vector.  The experimental determination of the real part of this value enables the construction of mean momentum flow lines which are independent of the photon content. Thus the flow lines cannot be interpreted as `photon trajectories'.

To produce evidence for so-called Bohmian mechanics, one needs to repeat the momentum weak value experiments using atoms  or neutrons as a source.  Neutron experiments have already been carried out by Sponar {\em et al.}~\cite{sstd15}.
Our group is developing techniques to make such measurements on atomic systems~\cite{rfbh14}.

\section{Acknowledgements}
We would like to thank Bob Callaghan, Glen Dennis, Chris Dewdney, Pan Kaloyerou, Peter Holland and Peter Barker for helpful discussions.
The authors would also like to thank the Fetzer Franklin Fund of the John E. Fetzer Memorial Trust for their continued support.



\bibliography{myfile}{}
\bibliographystyle{plain}
\end{document}